# McCall's Area Transformation *versus* the Integrated Impact Indicator (*I3*)

Sir:

In a study entitled "Skewed Citation Distributions and Bias Factors: Solutions to two core problems with the journal impact factor," Mutz & Daniel (2012) propose (*i*) McCall's (1922) Area Transformation of the skewed citation distribution so that this data can be considered as normally distributed (Krus & Kennedy, 1977), and (*ii*) to control for different document types as a co-variate (Rubin, 1977). This approach provides an alternative to Leydesdorff & Bornmann's (2011) Integrated Impact Indicator (*I3*). As the authors note, the two approaches are akin.

Can something be said about the relative quality of the two approaches? To that end, I replicated the study of Mutz & Daniel for the 11 journals in the Subject Category "mathematical psychology" of the Web of Science, but using additionally *I3* on the basis of continuous quantiles (Leydesdorff & Bornmann, in press) and its variant *PR6* based on the six percentile rank classes distinguished by Bornmann & Mutz (2011) as follows: the top-1%, 95-99%, 90-95%, 75-90%, 50-75%, and bottom-50%.[1]

---

[1] McCall's Area Transformation can be automated as a macro using the Excel function *normsinv*(); *I3* and *PR6* values can be computed using the software isi2i3.exe available at http://www.leydesdorff.net/software/i3 .

**Table 1**: Rank correlation (Kendall's tau) between the different JIFs and other variables (N = 11 journals).

|  | SC JIF* | $JIF_z$* | $cJIF_z$* | I3 | PR6 | N Pub* (08+09) |
|---|---|---|---|---|---|---|
| $JIF_z$* | 0.64[+] | | | | | |
| $cJIF_z$* | 0.60 | 0.89[+] | | | | |
| I3 | 0.56 | 0.78[+] | 0.89[+] | | | |
| PR6 | 0.49 | 0.71[+] | 0.81[+] | 0.93[+] | | |
| N Pub (08+09)* | 0.42 | 0.64[+] | 0.75[+] | 0.86[+] | 0.93[+] | |
| Citations 2010* | 0.59 | 0.81[+] | 0.92[+] | 0.99[+] | 0.92[+] | 0.84[+] |

[+] $p < 0.01$; * Source: Mutz & Daniel (2012: 174, Table 4).[2]

In Table 1, the rankings based on *I3* and *PR6* are correlated with the other rankings used in Table 4 of Mutz & Daniel (2012: 174).[2] The rank correlation of *I3* and *PR6* with the *cJIFz* (that is, the *JIFz* corrected for differences between journals in the proportions of document types) is very high (0.89 and 0.81, respectively) and statistically significant ($p < 0.01$; $N = 11$). The two sets of indicators are highly correlated (about .70 to .90), but not redundant or identical (because the correlation is smaller than 1.0).

Note that *I3* correlates above 0.99 with the number of citations and 0.86 with the number of publications. These correlations are higher than the 0.84 correlation between the numbers of publications and citations. Leydesdorff & Bornmann (2011, p. 2138, Figure 5) considered these high correlations important because impacts—unlike average impacts—depend on both the numbers of publications and their respective citations.[3]

---

[2] Mutz & Daniel (2012: 174) tested for significance at the 5% level.
[3] As expected, *I3* correlates somewhat less with the *N* of publications than with the *N* of citations. This difference disappears in the case of *PR6* as an effect of the nonlinear binning in six classes.

**Table 2**: *JIF* values for the WoS Category (WC) "Psychology, Mathematical," for the year 2010, compared with %I3 and %PR6.[a]

| Journal | N Pub* | N Cit* | SC JIF* | JIFz* | cJIFz* | %I3 [a] | %PR6 [a] |
|---|---|---|---|---|---|---|---|
| *Behav Res Methods* | 275 | 619 | 2.25 | 0.68 [1] | 0.80 [2] | 21.33 [2] | 22.26 [2] |
| *Psychon B Rev* | 329 | 712 | 2.16 | 0.63 [2] | 0.82 [1] | 37.50 [1] | 33.95 [1] |
| *Educ Behav Stat* | 48 | 69 | 1.44 | -0.37 [7] | -0.33 [7] | 4.25 [8] | 3.92 [8] |
| *Psychometrika* | 97 | 128 | 1.32 | 0.25 [3] | 0.19 [3] | 7.15 [3] | 7.06 [4] |
| *Brit J Math Stat Psy* | 64 | 78 | 1.22 | -0.18 [5] | -0.12 [6] | 4.44 [7] | 4.79 [7] |
| *J Math Psychol* | 94 | 111 | 1.18 | -0.09 [4] | 0.00 [4] | 7.04 [4] | 6.90 [5] |
| *Appl Psych Meas* | 82 | 78 | 0.95 | -0.52 [9] | -0.37 [8] | 5.70 [6] | 5.67 [6] |
| *J Educ Meas* | 52 | 41 | 0.79 | -0.85 [10] | -0.61 [10] | 3.07 [9] | 2.78 [9] |
| *J Classif* | 37 | 29 | 0.78 | -0.41 [8] | -0.39 [9] | 1.77 [10] | 2.37 [10] |
| *Educ Psychol Meas* | 118 | 89 | 0.75 | -0.25 [6] | -0.10 [5] | 6.51 [5] | 7.99 [3] |
| *Appl Meas Educ* | 44 | 13 | 0.30 | 1.05 [11] | -1.00 [11] | 1.22 [11] | 2.32 [11] |

\* Source: Mutz & Daniel (2012:174, Table 3).
[a] %I3 and %PR6 are used in order to ease the comparison.

In Table 2 the values of the *%I3* and *%PR6* are added to Table 3 of Mutz & Daniel (2012: 174). Unlike the *JIF* and *JIFz*, the three indicators *cJIFz*, *I3*, and *PR6* reverse the order between the two top journals: *Behavior Research Methods* and *Psychonomic Bulletin & Review*. The latter journal has more publications (in 2008 and 2009) and more citations (in 2010) than the former, but has a lower *JIF* and *JIFz*. One paper in *Behavior Research Methods* can be considered as an outlier with 270 citations in 2010. However, both journals have three papers in the top-1% of the set of 11 journals. In sum, *cJIFz*, *I3*, and *PR6* correct for outliers. Thus, the area transformation itself is not sufficient, but the normalization for different document types is additionally needed (Moed, 2010).

Let me note that *I3* is defined at the level of articles and thus allows for aggregations other than in terms of journals; for example, in the case of the evaluation of institutes or countries (Leydesdorff, 2011). McCall's (1922) Area Transformation and Rubin's (1977) Causal Model (that is, using document types as a co-variate) require other normalizations for differently aggregated document sets.

**Conclusions**

1. The results of Mutz & Daniel (2012) could independently be replicated;
2. In this sample (of 11 journals) the two sets of indicators are highly correlated. However, the correlations are far from perfect;
3. *cJIFz, I3,* and *PR6* correct for outliers while document types are taken into account.

Loet Leydesdorff

University of Amsterdam, Amsterdam School of Communication Research (ASCoR),

Kloveniersburgwal 48, 1012 CX Amsterdam, The Netherlands; loet@leydesdorff.net .

**Acknowledgement**

I thank an anonymous referee for his/her suggestions.